\documentclass{aastex631}

\usepackage{booktabs}
\usepackage{soul}

\newcommand{\ang}[1]{\ensuremath{{#1}^{\circ}}}

\newcommand{\is}{\textit{in situ}}
\newcommand{\eg}{\textit{e.g.}}
\newcommand{\kmps}{km.s${}^{-1}$}
\newcommand{\Br}{$B_{\rm R}$}
\newcommand{\Bt}{$B_{\rm T}$}
\newcommand{\Bn}{$B_{\rm N}$}
\newcommand{\Wind}{\textit{Wind}}

\newcommand{\ie}{\textit{i.e.}}

\shortauthors{Regnault et al.}
\graphicspath{{./}{figures/}}

\begin{document}

\title{Discrepancies in the Properties of a Coronal Mass Ejection on Scales of 0.03~au as Revealed by Simultaneous Measurements at Solar Orbiter and Wind: The 2021 November 3--5 Event}


\author[0000-0002-4017-8415]{F. Regnault}
\affiliation{Space Science Center, Institute for the Study of Earth, Oceans, and Space, University of New Hampshire}
\correspondingauthor{F. Regnault}
\email{florian.regnault@unh.edu}
\author[0000-0002-0973-2027]{N. Al-Haddad}
\affiliation{Space Science Center, Institute for the Study of Earth, Oceans, and Space, University of New Hampshire}
\author[0000-0002-1890-6156]{N. Lugaz}
\affiliation{Space Science Center, Institute for the Study of Earth, Oceans, and Space, University of New Hampshire}
\author[0000-0001-8780-0673]{C.~J. Farrugia}
\affiliation{Space Science Center, Institute for the Study of Earth, Oceans, and Space, University of New Hampshire}
\author[0000-0002-2917-5993]{W. Yu}
\affiliation{Space Science Center, Institute for the Study of Earth, Oceans, and Space, University of New Hampshire}
\author[0000-0002-5996-0693]{B. Zhuang}
\affiliation{Space Science Center, Institute for the Study of Earth, Oceans, and Space, University of New Hampshire}
\author[0000-0001-9992-8471]{E.~E. Davies}
\affiliation{Austrian Space Weather Office, GeoSphere Austria, Graz, Austria}

\begin{abstract}
%

Simultaneous \is\ measurements of coronal mass ejections (CMEs), including both plasma and magnetic field, by two spacecraft in radial alignment have been extremely rare. Here, we report on one such CME measured by Solar Orbiter (SolO) and \Wind\ on 2021 November 3--5, while the spacecraft were radially separated by a heliocentric distance of 0.13 au and angularly by only \ang{2.2}.
We focus on the magnetic cloud (MC) part of the CME. We find notable changes in the R and N magnetic field components and in the speed profiles inside the MC between SolO and \Wind. We observe a greater speed at the spacecraft further away from the Sun without any clear compression signatures. Since spacecraft are close to each other and computing fast magnetosonic wave speed inside the MC we rule out temporal evolution as the reason on the observed differences suggesting that spatial variations over \ang{2.2} of the MC structure are at the heart of the observed discrepancies.
Moreover, using shock properties at SolO, we forecast an arrival time 2h30 too late for a shock that is just 5h31 away hours from Wind. Predicting the north-south component of the magnetic field at \Wind\ from SolO measurements leads to a  relative error of 55\%. These results show that even angular separations as low as \ang{2.2} (or 0.03 au in arclength) between spacecraft can have a large impact on the observed CME properties, rising up the issue of the resolutions of current CME models and potentially affecting our forecasting capabilities.

\end{abstract}

\keywords{}

\section{Introduction}

The launch of Solar Orbiter (SolO, \citealt{muller2020}) and of Parker Solar Probe (PSP, \citealt{fox2016,raouafi2023}) enable measurements of coronal mass ejections (CMEs) in the innermost heliosphere at heliocentric distances below 0.5~au. Combined with older missions around 1~au, such as \textit{Wind}, Advanced Composition Explorer \citep[ACE,][]{chiu1998}, Deep Space Climate Observatory \citep[DSCOVR,][]{burt2012}, and Solar Terrestrial Relations Observatory \citep[STEREO-A/B,][]{kaiser2007,kaiser2008}, these measurements provide opportunities to study CMEs using multiple spacecraft distributed at different distances from the Sun. Such multi-spacecraft studies have so far focused on two main approaches depending on the distance between the spacecraft. The first is combining simultaneous measurements by two spacecraft within the CME , \ie\  at the same time but at different locations, often with moderate (typically $\sim 10--20^\circ$) angular separations in longitude, to investigate the three-dimensional nature of CME properties, as done, for example, by \citet{winslow2021}. This particular study relied on measurements from STEREO-A and PSP near 1~au, a location where PSP does not typically provide full plasma measurements. Because SolO performed Earth flybys, the mission has been making solar wind plasma and interplanetary magnetic field (IMF) measurements close to the Sun-Earth line and at distances of 0.5--1~au. The second approach relies on combining distant measurements from two spacecraft in approximate radial alignment to investigate CME propagation and evolution \citep[e.g., see][]{davies2021a,kilpua2021a,mostl2022} for CMEs first measured by SolO at a heliocentric distance of 0.8~au. Other studies before the launches of PSP and SolO used measurements in the inner heliosphere from planetary missions such as MESSENGER and VEX \citep[see, for example][]{winslow2015,good2016,salman2020a,lugaz2022}, which do not have plasma measurements. Earlier work used measurements from Helios \citep[e.g., see][]{leitner2007,delucas2011} in the 1980s but data quality and the lack of remote observations made reaching global conclusions more complicated. Regarding recent work with SolO,  \citet{davies2021a} studied a CME measured on 2020 April 19--21 by SolO, BepiColombo and Wind while SolO was at 0.81~au,  but no plasma measurements were available at SolO because the SWA suite commissioning was still ongoing. There was notable evolution of the CME between SolO and \textit{Wind}, with a decrease of the magnetic field strength in the flux rope by $\sim 35\%$ and an increase in duration by 10--15\%. Importantly, measurements of the flux rope at \textit{Wind} (on Apr. 20 07:56 UT) started just one hour before the end of the flux rope at SolO (Apr. 20 09:15 UT). Thus, a small portion of these measurements were made simultaneously at  two spacecraft locations and any discrepancies found may be thought more as the temporal evolution of the CME during the time it takes to pass over SolO rather than the effect of CME propagation from SolO to Wind.



In this article, we focus on a CME measured by SolO and \textit{Wind} on 2021 November 3--5 while they were separated by 0.13~au in heliocentric distance (\Wind\ was at 0.98~au and SolO at 0.85~au from the Sun) and both spacecraft provided plasma and magnetic field measurements. Their total angular separation was \ang{2.2} (\ang{0.95} in longitude and \ang{1.98} in latitude) corresponding to less than 0.04~au in arc length at 0.9~au. We compute it using the dot product of the two spacecraft position vectors. The CME is fast enough to drive a sheath with a fast forward shock at its front. Shock and sheath properties were analyzed recently by \citet{trotta2023}. In particular, they studied the shocklets associated with the sheath region observed by \textit{Wind} and, interestingly, they did not observe shocklet signatures at SolO. Their analysis did not extend to the magnetic cloud (MC) region of the CME. The latter corresponded to the magnetic structure being expelled from the corona where we observe its specific \is\ signatures detailed in Section \ref{sec:MC_boundaries}. Moreover, they did not make use of the fact that for a time, SolO and \textit{Wind} were probing different portions of the MC at the same time. It is this ``gap'' which we wish to bridge in the present work. With multiple spacecraft probing the MC at different locations at the same time, one can determine the instantaneous properties of the MC by combining measurements taken by both spacecraft \citep{regnault2023b}.

The layout of the paper is as follows.
In Section \ref{sec:context}, the CME properties when close to the Sun and the heliospheric context in which the CME propagated are introduced. In Section \ref{sec:compare}, we compare the properties of the magnetic field profile inside the MC at SolO and at \textit{Wind}. In Section \ref{sec:inst} we combine both profiles to obtain the instantaneous properties of the CME. In Section \ref{sec:forecast} we compare what would have been the forecasted properties at \textit{Wind} given those at SolO in the context of space weather with actual measurements at \textit{Wind}. Finally, in Section \ref{sec:concl}, we discuss our findings and draw our conclusions.


%

\section{Overview of the In Situ Measurements for the 2021 November 3--5 CME}
\label{sec:context}




\begin{figure*}[ht!]
    \centering
    \includegraphics[width=.9\textwidth]{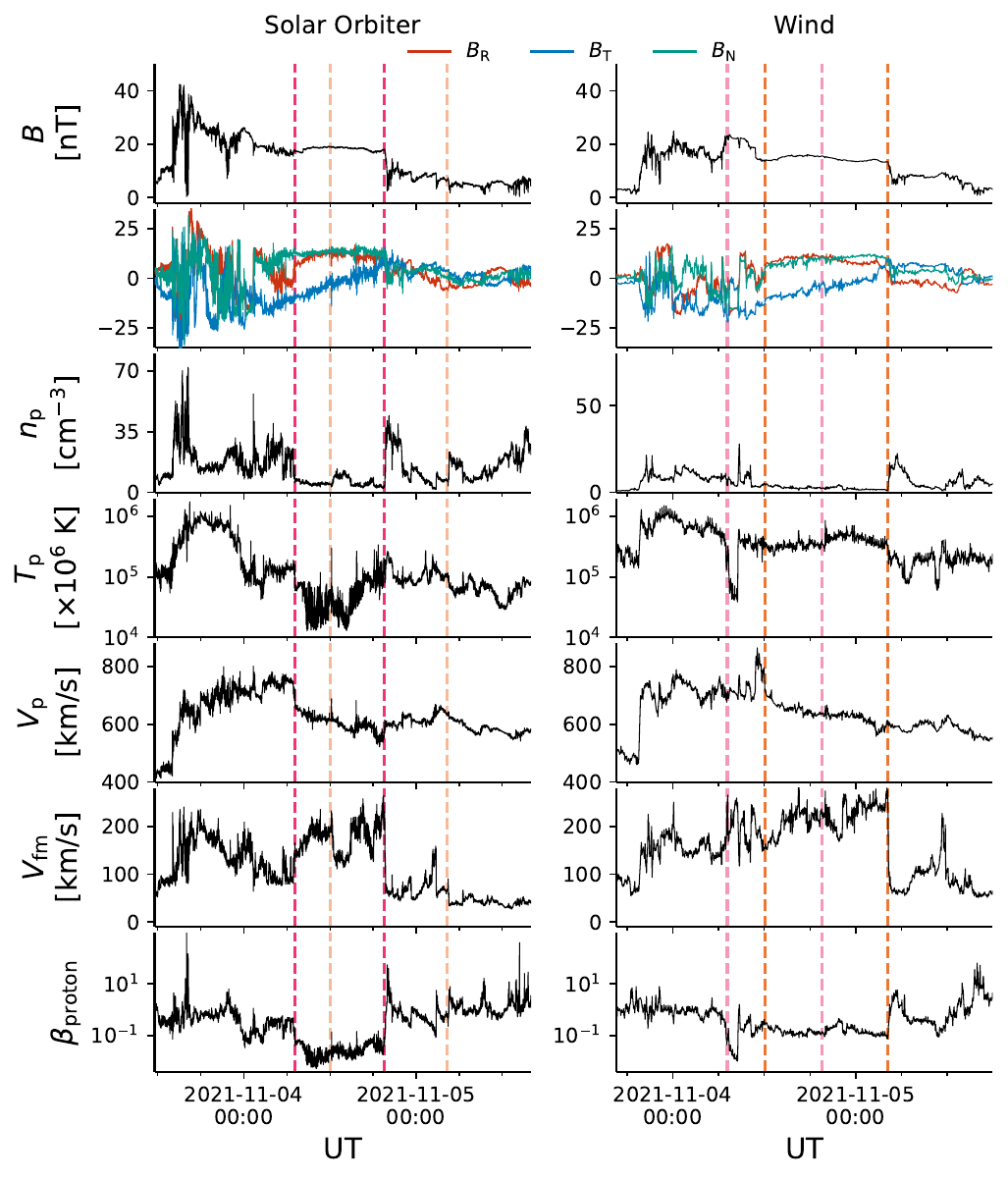}
    \caption{Magnetic and plasma measurements at Solar Orbiter (left) and \textit{Wind} (right) during the passage of the 2021 November 3--5 CME. From top to bottom: the magnitude of the magnetic field (in nT), its components (in nT) in RTN coordinates, the proton density (in cm${}^{-3}$), the temperature (in $\times 10^{6}$ K), the bulk speed (in \kmps), the fast magnetosonic speed (in \kmps) and the proton $\beta$. MC boundaries are delineated in magenta for SolO and orange for Wind. The vertical axis of the SolO and \textit{Wind} panels are on the same scale.}
    
    \label{fig:insitu}
\end{figure*}

\subsection{Magnetic Cloud Boundaries and CME Coronal Counterpart}
\label{sec:MC_boundaries}

The CME discussed in this article was measured {\it in situ} starting on 2021 November 3.
Figure~\ref{fig:insitu} from top to bottom shows the magnetic and plasma (density, temperature and speed of the protons) measurements along with the fast magnetosonic speed (described later in this section) and the proton-$\beta$ (thermal pressure over magnetic pressure, \citealt{richardson1995,wang2005}) measured by Solar Orbiter (left) and \textit{Wind} (right) at the time of the passage of the CME. Respectively, for the magnetic and plasma measurements at SolO, we use the MAG \citep{horbury2020} and SWA \citep{owen2020} instruments and for \textit{Wind} we use the MFI \citep{lepping1995} and SWE \citep{ogilvie1995} instruments.

The present work focuses on the study of the MC part of the CME. MCs are characterized by (i) an above average magnetic field strength; (ii) a smooth and large rotation of the magnetic field vector, and (iii)  a low proton-$\beta$ \citep{burlaga1981}.
However, using only these criteria, the MC start time at \textit{Wind} would be hard to define. We thus decide to start the MC where the speed profile starts to decrease monotonically following a linear trend, as expected for an expanding MC  \citep{burlaga1982,farrugia1993,gulisano2010}. 
The end of the MC is chosen to be when the proton $\beta$ increases suddenly along with a drop in the magnetic field strength at both spacecraft. A more detailed description of the MC properties can be found in Section \ref{sec:desc}.
Using these boundaries, the MC starts on 2021-11-04T07:09 UT and ends on 2021-11-04T19:38 UT at SolO. For Wind, the MC starts on 2021-11-04T12:06 UT and ends on 2021-11-05T04:14 UT.
MC boundaries are shown in magenta for SolO and orange for Wind. To show the overlap of the \is\ measurements, MC boundaries at \textit{Wind} appear in a lighter color in the SolO panel (left), and vice versa. 


The coronal counterpart of this event should have erupted between October 30 and November 1 since a CME typically takes 2 to 4 days to reach the Earth \citep{gopalswamy2001}. However, in this time interval, 5 CMEs are observed in both STEREO/COR and SOHO/LASCO coronographs. \cite{li2022} studied these CMEs and found, by using the Graduated Cylindrical Shell model \citep[GCS, see][]{thernisien2011} that only one CME (their CME8) was directed toward Earth. We thus consider that it is this CME of interest that impacted SolO and \textit{Wind}.
%
This CME propagated with a direction of \ang{20}N and \ang{9}E with respect to the Sun-Earth line, with a tilt angle of \ang{-49}. The result of the reconstruction technique can be seen in Figure 2 of \citet{li2022}. We performed the fitting individually with the same reconstruction model and found the same result of the propagation direction within \ang{2}. Given this direction and assuming a self-similar expansion along with a radial propagation, we would expect the CME flank to cross over SolO and \textit{Wind}.

We inspect EUV and coronagraphic images to determine if there were low-latitude coronal holes or other CMEs launched after the eruption of the CME presented in this study that could cause deflections or impact its propagation \citep{gopalswamy2009b,scolini2021g}. We found no evidence of any.
The CME under study is therefore expected to propagate undisturbed and radially in a relatively ``pristine'' solar wind.







\subsection{Description of the MC In Situ Properties}
\label{sec:desc}

Using the average speed of the MC and its duration, we find that the MC has a radial width of 0.18~au at SolO and 0.25~au at Wind. \cite{liu2005} fitted the evolution of the observed size of CMEs as a function of distance and found the following propagation law: $S(R) = (0.25 \pm 0.01) \times R^{0.92 \pm 0.07}$ with $R$ the distance from the Sun in au. With the size at SolO, we then find that the MC at \textit{Wind} should have a size of 0.21 au which is smaller than the size we actually measure. This shows that this statistical relationship does not explain the increase in the MC radial size of this event from SolO to Wind. It is important to note that, to the best of our knowledge, the 0.92 power law index from \cite{liu2005} is one of the highest that we can find in the current literature. Then 0.21 au correspond to a MC size estimate at Wind from SolO measurements on the higher end. This could be due to the different trajectory of spacecraft through the MC even though they are close to each other. This matter is discussed in more details in the rest of the paper.

The sixth row of Figure \ref{fig:insitu} shows the fast magnetosonic speed $V_{\rm fm}$ at SolO and at \textit{Wind}. It corresponds to the maximum speed that fast magnetosonic waves propagate in a magnetized plasma defined as follows:
\begin{equation}
v_{\rm fm}^2 = v_a^2 + c_s^2, 
\end{equation}
where $v_a$ is the Alfv\'en speed and $c_s$ is the speed of sound. 
We take $\gamma = 5/3$ to derive the sound speed.

Quantity $v_{\rm fm}$ corresponds to the fastest way for information to propagate from one point in the MC to another. 
We compute its average value in the MC at both spacecraft and find 175 \kmps\ at SolO and 219 \kmps\ at \textit{Wind}. We find here that $v_{\rm fm}$ increases from SolO to \textit{Wind} which is not what we expect since \textit{Wind} is further from the Sun and $v_{\rm fm}$ is expected to decrease with distance in the solar wind. This increase is due to the fact that the average density of the MC at \textit{Wind} is much lower than at SolO (5.7 vs 2.6 cm${}^{-3}$, respectively) while the average magnetic field is similar (18.2 vs 14.5 nT).

The shock time observation is 14:04 UT at SolO and 19:35 at Wind on the 2021 November 3 \citep{trotta2023} which leads to a time delay of  5 hours and 31 mins between SolO and Wind. Within that time, fast magnetosonic waves have time to travel between 0.02 to 0.03 au (depending on whether the speed based on SolO or \textit{Wind} measurements are used), which is much less than half of the estimated MC size. We use the shock arrival time to determine the delay of the CME between SolO and \textit{Wind} rather than the MC boundaries because the shock is the easiest feature to locate and thus give the most accurate time delay estimate.

These considerations will be important when determining whether the discrepancies in the magnetic field highlighted in Section \ref{sec:compare_mag} can be explained only with time evolution or if other reasons, such as the angular separation between the spacecraft, need to be invoked.


\section{Comparing the MC properties}
\label{sec:compare}

In this section, we compare the velocity and magnetic field measurements made by the two spacecraft in the MC region to see how different they are at two spacecraft that have an angular and radial separation of only \ang{2.2} and 0.13~au.
In order to establish how much of any discrepancy can be attributed to instrumental effects, we compute the average difference between magnetic field and speed measurements made at \textit{Wind} and SolO for small separations. This is presented in Appendix \ref{sec:cross-correl}. To the best of our knowledge, such a cross-calibration of SolO with L1 measurements has not been published. In summary, when SolO and \textit{Wind} are separated by less than 0.01~au, the magnetic field measurements (the magnitude, and $T$ and $N$ components) differ (on average in absolute value) by 0.4~nT, the $B_R$ component by 0.5~nT and the speed by  5.18~\kmps. 
Therefore, any difference in the speed or in the magnetic field that significantly exceeds these values that may be found during this event cannot be fully explained by instrumental effects.

\subsection{Magnetic Field Time Profile}
\label{sec:compare_mag}

\begin{figure*}[ht!]
    \centering
    \includegraphics[width=.9\textwidth]{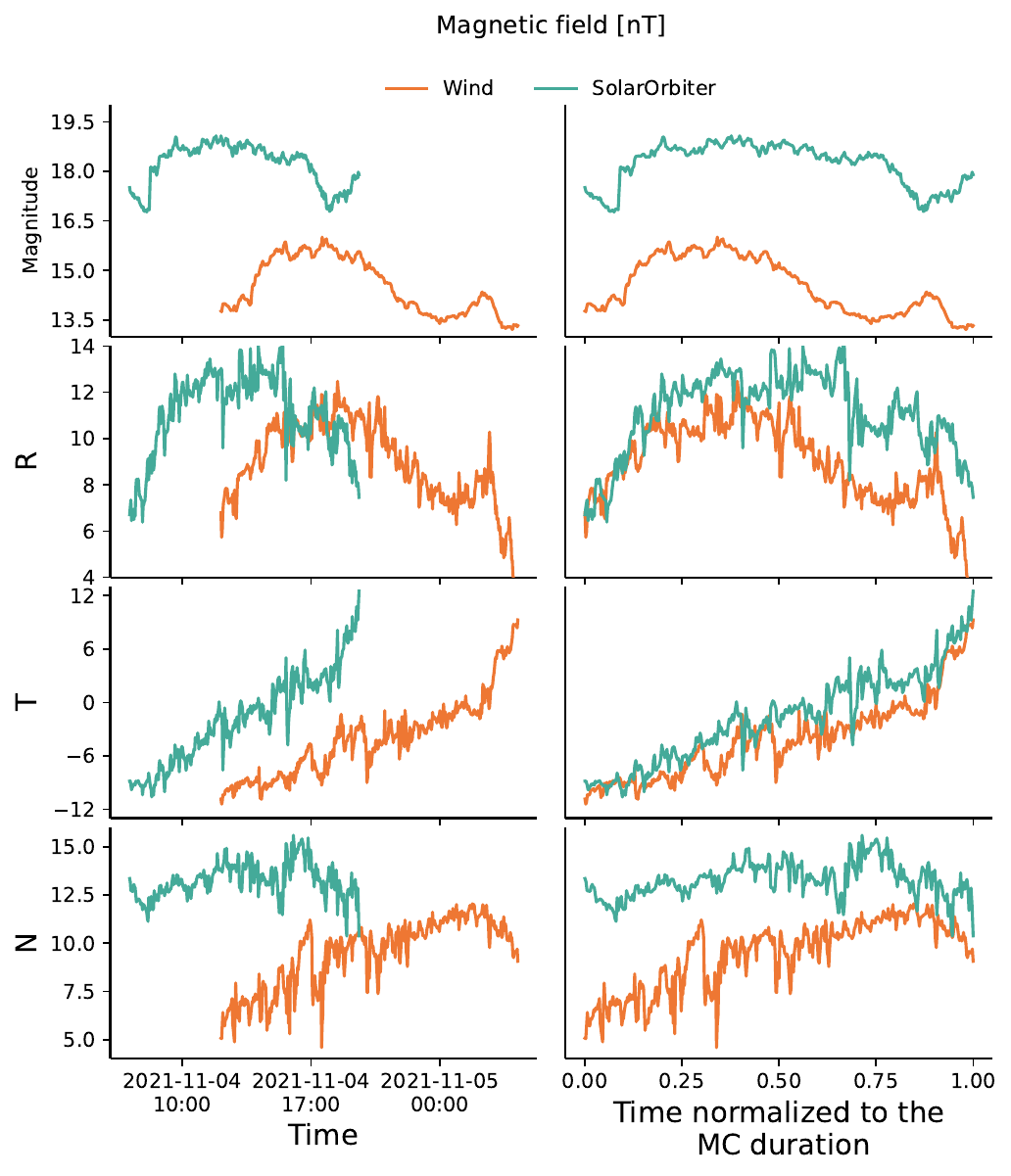}
    \caption{Magnetic field (in nT) strength and its components in RTN coordinates at \Wind\ (orange) and SolO (green) as a function of time (left panel) and time normalized to the MC duration (right panel).}
    \label{fig:overplot}
\end{figure*}

Figure~\ref{fig:overplot} shows the magnetic field and the speed and their respective components measured by \textit{Wind} and at SolO. 
The left column of Figure \ref{fig:overplot} shows the magnetic field strength and its components in Radial-Tangential-Normal (RTN) coordinates when no time shift is applied to the data. The right column shows the same parameter but plotted as a function of the time normalized to the MC duration. The left column highlights the strong time overlap between the profiles measured at SolO and \Wind\ while the right column put emphasis on the differences in the components of the magnetic field profiles measured by both spacecraft that we describe now.

A stronger average magnetic field strength is measured at SolO than at \textit{Wind}, which is somewhat expected due to the expansion of the MC during its propagation from SolO to \textit{Wind}. The average magnetic field in the MC is 18.2~nT at SolO and 14.5~nT at \textit{Wind}. The evolution of the magnetic field as a function of the distance from the Sun is often described by a power law ($B(r) \propto r^{-\alpha}$ with $\alpha = 1.3 \textendash 1.8$, \citealt{leitner2007,winslow2015,davies2021b}). Using the two positions of the spacecraft and the average magnetic field at both spacecraft we find that $\alpha \sim 1.6$ which fits within the observation range of previous statistical studies. Both magnetic field strength profiles are approximately flat, in particular when looking with the $y$-scale used in Figure \ref{fig:insitu}.

\Br\ shows a similar shape at both spacecraft. However, while both \Br\ profiles start at the same value at the MC front, they start to differ measurably close to the MC center. At the MC rear, they differ by about $\sim 4$~nT.  A smooth rotation of $B_{\rm T}$ is measured at both spacecraft, as expected for two spacecraft with such low angular separations \citep{lugaz2018}. Under the flux rope paradigm, the smooth rotation along the $T$ direction suggests a highly inclined flux rope (with the axis $\sim$ perpendicular to the ecliptic plane). We also find that this component of the magnetic field is similar at both spacecraft, with a slight shift towards lower values at \Wind. 
\Bn\ is the component that shows the strongest discrepancies when comparing measurements made by both spacecraft. While SolO measures a \Bn\ approximately constant around 12.5~nT, \Bn\ is found to start at about 6~nT and almost constantly increasing at \Wind, reaching about 10~nT close to the MC rear. This is a difference of about 50\% in \Bn\ for two spacecraft separated by less than 0.13~au. Assuming the magnetic field components also follow a power law, this decrease of \Bn\ corresponds to a power law index $\alpha_{\rm N} > 4$, which is highly unrealistic given constraints on magnetic field conservation and evolution. This is further discussed below.

Discrepancies discussed above can be interpreted as a physical phenomenon occurring within $\sim$ 5 hours and 31 minutes (time delay between the shock observations at the two spacecraft) which can significantly change the magnetic field properties, such as magnetic reconnection or relaxation of the MC after being compressed at the rear. This point is discussed in further detail in Section \ref{sec:origin}. 
It can also be interpreted as the effect of the, although small, angular separation between the spacecraft. It would suggest that \ang{2.2} is enough of a separation to result in significant changes in the magnetic field profiles. \cite{lugaz2018} studied the correlation of the magnetic field strength profile observed by ACE and \textit{Wind} while separated by different angles around 0.5--1$^\circ$. Extrapolating their results, they found that the magnetic strength profile correlation between the two spacecraft reaches 0 when the angular separation becomes as large as \ang{15}--\ang{20}. We emphasize that an angular separation of \ang{5}--\ang{10} is often considered as a ``good" radial alignment \citep{good2019,salman2020a}, but we show here that significant discrepancies appear even for separations as small as $\sim \ang{2}$. As discussed further below, under the assumption that MCs are twisted flux ropes with an axial invariance, most of the angular separation between SolO and \textit{Wind} occurs along the flux rope axis, \ie, the difference between SolO and \textit{Wind} measurements are expected to be extremely small unless there is no axial invariance of the magnetic field over scales of \ang{2}.

\subsection{Velocity Time Profile}
\label{sec:compare_speed}
 
We now address the velocity differences, as highlighted in the fifth panel of Figure~\ref{fig:insitu} to compare the velocity profiles at SolO and \textit{Wind}. The MC is faster at \textit{Wind} than at SolO, which is somewhat unexpected. The center speed of the MC is 637 \kmps\ at \textit{Wind} and 604 \kmps\ at SolO. If both spacecraft probe a similar portion of the CME, which is expected to be true for small angular separations, this suggests an acceleration of the CME from SolO to \textit{Wind}. This is not expected because CMEs that are faster than the solar wind slow down as they propagate due to the drag caused by the solar wind \citep{vrsnak2013}.
In this case, the solar wind before the CME at SolO is slower by $\sim 150$~\kmps\ than its MC center speed, which indicates that the CME should decelerate instead of accelerate, based on our current understanding of CME-solar wind interaction. At \textit{Wind}, the solar wind before the CME is slower by $\sim 120$~\kmps. Therefore, the apparent acceleration cannot be explained by currently existing models of CME propagation and CME-solar wind interaction, and another explanation, such as the effect of the angular separation, needs to be invoked.

The $T$ and $N$ components of the velocity (not shown) are close to 0 \kmps\ on average ($<10$ \kmps) at both spacecraft. Thus the $V_{\rm R}$ ($\sim V_{\rm mag}$) profile shows a typical almost monotonic decrease at both spacecraft. 
These measurements thus suggest that the bulk motion of the MC is locally approximately radial, at least during the duration of the passage of the CME over the spacecraft. 
The lack of non-radial flows and its consequence for CME expansion were recently discussed in \citet{al-haddad2022}, and the measurements here are consistent with the findings from this study. Overall, non-radial propagation and complex expansion (which would result in non-radial flows) cannot be invoked to explain the differences between the two spacecraft.

In addition, the bulk motion of the plasma deduced from the speed profile is not consistent with the time delay in the CME arrival at SolO and Wind. The time delay of the MC front between \textit{Wind} and SolO is only about 4 hours. However, if the MC front does propagate radially, based on the speed measured and the separations between the spacecraft, it would take more than 8 hours to travel 0.13~au (heliocentric distance separation between SolO and Wind) at about 650 \kmps. The differences in the speed as well as the \Bn\ component of the magnetic field inside the MC point towards spatial differences in the MC properties due to the small angular separation between Solo and \Wind. We discuss this further in the next section.


%
%
%

\subsection{Regarding the Origin of the Discrepancies in the Measurements between Solar Orbiter and Wind}
\label{sec:origin}

In Sections \ref{sec:compare_mag} and \ref{sec:compare_speed}, we showed that \Wind\ and SolO measured different MC magnetic and speed properties. In particular, the magnetic field components (specifically \Bn) suggest a change in the MC structure and the speed profiles suggests an acceleration of the MC.
A possible explanation for these discrepancies could be an interaction with the solar wind \citep[e.g., as discussed in][]{scolini2021g}. For instance, a fast solar wind stream could catch up the MC rear, compress it and could also result in a bulk acceleration of the MC. However, no compression signatures are observed at the rear of the MC, such as increased magnetic field strength, density and temperature. Moreover, a MC being compressed from the rear would typically show an asymmetric magnetic field profile at SolO or \Wind\ which is not observed. While there is a small rise of the speed close to the rear of the MC, a fairly monotonic decrease of the speed is measured at both spacecraft. If a solar wind stream was pushing the MC from behind, the $V_{\rm mag}$ profile would exhibit a gradient reversal. The speed profile after the end of the MC  shows a continuation of the decrease at \textit{Wind} (see Figure~\ref{fig:insitu}), not consistent with a high-speed stream behind the MC.

For this event, there is an increase of the density up to 40 cm${}^{-3}$ at SolO and 20 cm${}^{-3}$ at \textit{Wind} compared with the 2--4~cm${}^{-3}$ inside the MC (see Figure~\ref{fig:insitu}). However, we do not observe significantly faster flows than the MC center speed nor heated plasma at the rear of the MC. We thus do not find any clear signs of compression of the plasma. 

In addition, the very short time ($\sim$ 5.5 hours) between the MC observations allows us to rule out time evolution as the main contributor to the observed discrepancies. Indeed, according to the speed of fast magnetosonic waves described in Section \ref{sec:desc}
, waves only have time to travel 0.03~au of the MC as we discussed in Section \ref{sec:context}.
However, it is clear from the data shown in Figure \ref{fig:overplot} that discrepancies in the magnetic field strength profile between the MC at SolO and at \textit{Wind} are not just contained in the last 0.03~au but rather in about the last half of the MC for \Br\ and for almost all the MC for \Bn\ which corresponds respectively to a length of 0.09~au and 0.18~au at SolO, respectively.



Another candidate to explain the differences would be reconnection occurring in the MC. However, this process is often very local and thus is unlikely to explain the discrepancy over such a large part of the MC (half of its radial size).  Moreover, we do not observe any reconnection signatures (such as jets of fast plasma and/or enhanced temperatures, or pairs of shocks) in the MC \citep{owens2009,xu2011}.

To conclude, we have shown in this section that the most commonly advanced reasons to explain the discrepancy in the properties of the MC at \textit{Wind} and at SolO are not consistent with the timing of the CME at both observations, neither with the evolution of \Bn\ from SolO to \Wind. 
The type of diagnostic described in this section is made possible mainly because both spacecraft have a radial separation small enough that  fast magnetonosic waves do not have time to travel through the full MC part during the $\sim 5.5$ hours between the measurements.
Since time evolution fails to explain the discrepancies we observe, such effect is likely produced from the variations of the MC properties on scales as small as $\sim 0.03$ au which has almost never been explored with simultaneous measurements.


\section{Instantaneous Expansion Speed}
\label{sec:inst}

In Section \ref{sec:compare}, we have described measurable differences between the magnetic field and speed \is\ profiles measured at SolO and at \textit{Wind} for a MC measured at small angular separations and moderate radial separations. In this section, we perform a thought experiment to emphasize the impact of a \ang{2.2} angular separation on the MC properties. This approach involves demonstrating that the assumption under which both spacecraft observe the same portion of the CME leads to inconsistencies in the quantification of the MC expansion between the standard estimation using a single \is\ profile and the instantaneous expansion speed. The estimate of the instantaneous expansion speed is made possible thanks to the specific spacecraft configuration when the CME was observed. To the best of our knowledge, this is the first report of such estimate through \is\ measurements only.

MC expansion speed has often been defined using the following formula \cite[\eg][]{owens2005,gulisano2010,gopalswamy2015}:
\begin{equation}
    V_{\rm exp} = \frac{V_{\rm front} - V_{\rm back}}{2}
    \label{eq:vexp}
\end{equation}

This formula assumes that all the change in the MC speed profile is due to the expansion of the MC. That is to say, the MC does not significantly accelerate nor decelerate during the time the CME passes over the spacecraft. Under the same assumption, the expansion speed is sometimes defined using the slope of the speed profile during the part of the event for which the speed can be reasonably approximated by a linear function \citep[]{gulisano2010,lugaz2020a}. Again, any bulk acceleration or deceleration shall affect the calculation of the expansion speed.

We perform a linear fit to the MC speed profiles at both spacecraft through the entire MC duration as the speed is close to linearly decreasing.
Using the result of this fit, we find the expansion speed at \textit{Wind} and SolO to be 43 and 41~\kmps, respectively, which is a consistent value between the two spacecraft and also a typical value found in the literature: \citet{gopalswamy2015} found an average expansion speed for solar cycles 23 and 24 of 51 \kmps\ and 25 \kmps, respectively. With an expansion speed of 43 \kmps, the MC size that was 0.18~au at SolO should increase by 0.007~au when it reaches \textit{Wind}, which would result in a significantly smaller size than the 0.25~au we calculate from {\it in situ} measurements at \textit{Wind}. We thus find that the measured expansion at SolO cannot explain the increase in size from SolO to \textit{Wind}. One reason could be that the assumption that the MC speed is constant during the time the CME passes the spacecraft is not justified. 

Using the specific spacecraft configuration during the passage of the MC, we can estimate an instantaneous expansion speed. 
When SolO is probing the center of the MC, \textit{Wind} is probing its front, as the two spacecraft radial separation of 0.13~au is approximately equal to half the MC size at \textit{Wind} (see also left panel of Figure~\ref{fig:overplot}). We thus have a measurement of the speed at the MC front and at its center \emph{at the same time.} We are then able to compute an instantaneous expansion speed combining measurements made by SolO and by \textit{Wind} with the following formula:

\begin{equation}
    V_{\rm exp_{\rm inst}} = V_{\rm front_{\rm Wind}} - V_{\rm center_{\rm SolO}}
\end{equation}

\begin{figure*}[ht!]
    \centering
    \includegraphics[width=.8\textwidth]{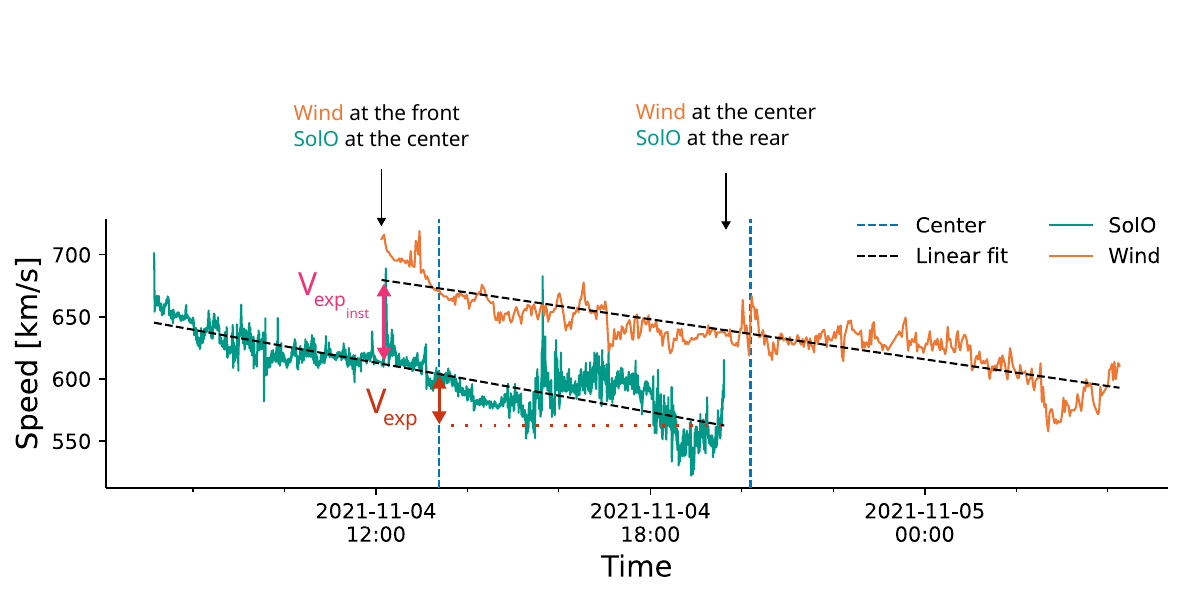}
    \caption{Schematic of the measurement of the instantaneous expansion speed.}
    \label{fig:inst_exp}
\end{figure*}

This is summarized in Figure \ref{fig:inst_exp} that shows the speed profile at SolO and \textit{Wind} in green and orange, respectively. No time shifting is applied to highlight the instantaneous nature of the measurements at SolO and \textit{Wind} for about half of the MC duration. Blue dashed lines show the center in time of the MC at both spacecraft. 
The center speed is assumed to be the speed at the central time of the fit, defined as the cruise (or bulk) speed in \cite{owens2005}.


We now compute the instantaneous expansion speed $V_{\rm exp_{\rm inst}}$. We use both linear fits of the speed and take the difference during the overlapping time to obtain a value that starts at 68 and increases up to 78~\kmps.
On average, we thus obtain $\bar{V}_{\rm exp_{\rm inst}} = 73$~\kmps, which is 1.7~times the expansion speed calculated using a single speed profile in the MC. This gap is greater than the average absolute difference between the measurements made of the bulk velocity described in Section \ref{sec:cross-correl} (5.18~\kmps), and is thus significant. 
Using this instantaneous expansion speed, the increase in size from SolO to \textit{Wind} would then be 1.7~times the one we find from a single velocity profile \ie\ 0.012~au. This leads to a predicted size at Wind of 0.192~au which is still significantly smaller than the 0.25~au that is measured at this location.

Let us now consider the broader scope of the global study rather than focusing solely on the thought experiment conducted in this section. This result could indicate that the MC is accelerating when passing over SolO and \textit{Wind} which flattens the ``normal" speed profiles, reduces its slope and thus decreases the expansion speed we infer from them. This is also consistent with the increased center speed observed at SolO and Wind. However, in the absence of clear compression signatures (see Section \ref{sec:origin}) of the MC at both spacecraft or a high speed stream behind the MC, the origin of such acceleration at about 0.8--1 au remains mysterious, and spatial (rather than temporal) variations within the MC, even at scales of \ang{2.2}, are a more likely explanation.
In addition, this increase of the expansion speed is not enough to explain the large increase of the size from SolO to \textit{Wind} highlighted in Section \ref{sec:desc}. 

This explains why we interpret this result as the fact that the assumption under which both spacecraft probe the same portion of the MC does not hold even when spacecraft are within \ang{2.2} from each other. To summarize, results presented in this section highlight how impactful a \ang{2.2} angular separation between two spacecraft can be on MC properties deduced from \is\ profiles.

\section{Impact of the Angular Separation on the Forecasted Properties at 1 au}
\label{sec:forecast}

In the context of space weather analysis, SolO was in a very good position (\ie\ an almost perfect radial alignment and upstream of L1) to perform forecasting of the CME properties or its arrival time at the Earth. It is this exercise that we perform in this section. Note, however, that we use science-level data from SolO, which differ from the real-time data streams, so this is an exercise of hindcasting rather than forecasting.

The shock arrival time is computed using the speed right after the shock as we can observe in Figure \ref{fig:insitu}. At SolO, the front of the sheath has a speed of about 680 \kmps. Thus, the shock can be forecasted to take 7 hours and 58 minutes to travel the 0.13~au separating SolO and \textit{Wind}. In reality, it took only about 5 hours and 31 minutes. Such an estimate would then have provided an arrival time that was 2.5 hours too late when the MC was just 5.5 hours away. It is a really notable error given the short distance between SolO and the Earth when the CME was observed. It corresponds to a relative error on the arrival time of 45\%.

In addition, the CME speed would be underestimated by at least $40$ \kmps\ and even more if the effect of the drag on the CME speed is taken into account. In order to have a rough estimate of how much the CME would have decelerated from the heliocentric distance of SolO to that of \Wind\ due to the drag effect, we use the Drag Based Model \citep[DBM,][]{vrsnak2013}. This model is often used to estimate CME arrival time and impact speed. As inputs, the CME coronagraphic speed of 1235 \kmps\ as found in \cite{li2022} and an asymptotic solar wind speed of 450 \kmps\ as measured in Section \ref{fig:insitu} before the CME arrival time are used. The drag parameter $\Gamma$ is determined to obtain a CME speed of about 670 \kmps\ at SolO location (corresponding to the observed MC front speed). We find that a drag parameter of $0.24~ 10^{-7}$ \kmps\ produces an impact speed of 669 \kmps. This $\Gamma$ value fits within its typical range described in \cite{vrsnak2013}. According to this model, the CME decelerates by about 30 \kmps\ from SolO to \textit{Wind} due to the drag. Combining the measured speed increase with the expected speed decrease, the total underestimation using SolO measurements is of about 70 \kmps for the CME front speed.

We note that with this $\Gamma$ value, the DBM model predicts a CME arrival time on 2021-11-03T15:00 which is a mismatch of more than 16 hours with the actual observations and increasing $\Gamma$ to reduce this mismatch leads to a CME speed of about 500 \kmps\ at SolO which is much lower to the one measured.


On top of the arrival time and speed estimates, the prediction of the geoeffectivity of the CME could have been significantly off too. Indeed, the north-south component of the magnetic field ($\sim B_{\rm N}$ ) is the most important for determining the geoeffectivity of a CME. A large and prolonged southward-oriented magnetic field is the most geoeffective magnetic field configuration \citep[]{Gonzalez1987}. While the $B_N$ component is mostly positive (northward magnetic field) during this MC, we discuss the difference between the two spacecraft under the assumption that such differences would have applied, even if $B_N$ had been negative.

While the magnetic field strength of the MC front at both spacecraft matches well, the $B_{\rm N}$ values differ by about 6~nT ($\sim$ 12.5~nT at SolO and $\sim$ 6.5~nT at Wind). If we take into account the expansion of the magnetic field from SolO to \textit{Wind} and with $\gamma = 1.8$ we would expect a decrease of $B_N$ from SolO to Wind of 2.4~nT, and there is still an unexplained different of 3.6~nT. This is a 55\% relative error on the geoeffective component of the magnetic field of MC that we would have forecasted at the Earth.
Again, since the CME had a northward magnetic field within the MC, it was not likely to be geoeffective anyway, but such relatively large differences in the $B_N$ component of the magnetic field from two spacecraft in close proximity is worrisome for space weather forecasting using upstream monitors.
To summarize, even if the MC was only 0.13~au away and with a \ang{2.2} angular separation, current models would have failed to accurately predict the speed, the arrival time at Wind and the magnitude of the \Bn\ component of the magnetic field at Earth, with errors of 40--50\% for lead times of $\sim$ 4.5 hours.

\section{Discussion and conclusion}
\label{sec:concl}

In this paper, we have shown that even when two spacecraft have an angular and radial separation of \ang{2.2} and 0.13~au, respectively, significant discrepancies in their magnetic and speed parameter profiles can exist. 
The angular separation is very small as compared to the typical angular extent of CMEs seen in coronographs (\ang{60}, \citealt{wang2011}). Such an angular separation corresponds to about 0.03~au in arclength at 0.9~au. Given this small separation, we initially expected both spacecraft to probe the same portion of the CME and thus observe similar features in the \is\ profiles. However, some of the magnetic field components are significantly different between the two spacecraft. While the expansion speed is the same at both spacecraft, speed profiles suggest an acceleration of the CME from SolO to \textit{Wind}. This is indicated both by the increase of the center speed from SolO to \textit{Wind} and with the flattening of the speed profile inferred by the comparison of the expansion speed with, and without, the effect of the CME temporal evolution. But it is not clear what physical process could cause such an acceleration rather than the expected deceleration due to the drag force exerted by the solar wind. 
We thus attribute these discrepancies to the effect of the variation of the MC across its cross-section within \ang{2.2} (or an arclentgh of 0.03~au), even though the separation is primarily in the direction of the expected MC axis under the current flux rope paradigm.



Over the last years, researchers have had more opportunities than before to perform multi-spacecraft studies of the same CME by spacecraft separated by 0.2--0.8~au and by 1--10 \ang{} while typically assuming that the angular separation has a smaller impact than radial separations. The approach thus assumes that the CME is coherent within these spatial scales.
The current work sheds light on the importance of ``small" spatial scales (\ie\ as small as $\sim 0.03$~au of arclength) on the MC properties. \cite{davies2020} found discrepancies in the results of fitting outputs of the same CME observed in particular by \Wind\ and Juno. At the time of the CME passage, \Wind\ and Juno were radially and angularly separated by 0.24~au and \ang{3.6} respectively. The time delay between the CME front at both spacecraft in their study was about 20 hours. In this study, we find similar discrepancies but with only about 5 hours of time delay (for the shock and MC front). This clearly highlights that ``small'' angular separations may have a significant impact on the MC properties as measured by two spacecraft. In fact, for this event, the effect of the small angular separation is clearly larger than the impact of the radial separations between the spacecraft, even though the separation is primarily along the MC axis (\ie\ there is limited influence of a different impact parameter). 

Moreover, trying to use the CME speed at SolO to hindcast the arrival time at Wind results in relative error of 45\%. According to \cite{riley2018} and \cite{wold2018} the typical error for the CME arrival deduced from its coronal properties (potentially coupled with simulations)  is 10 hours which corresponds to a relative error of 13\% taking 3 days as the typical CME transit time. Thus, our findings suggest that local measurements made just 0.13~au ahead and \ang{2.2} off the Sun-Earth line provide, in this case, a worse prediction than when global properties close to the Sun deduced from remote sensing are used. This sheds light once again on the effect of the local nature of the \is\ measurements and the need for more multi-spacecraft measurements of the same CME at small separations.

This study presents a case study of a CME with both magnetic and plasma signatures observed by two spacecraft simultaneously in the MC while separated by only \ang{2.2}. The event presented here is almost unique: when searching over more than 2000 CMEs observed over the last 40 years of data, it was the only event with such a spacecraft configuration along with clear CME signatures with both magnetic and plasma data available. The results found here cannot be generalized to all CMEs, but it raises important questions about the coherence scales of the CME magnetic structures. The coherence scale of CMEs have been estimated recently by \cite{lugaz2018,owen2020} using \is\ measurements and they found an angle ranging from \ang{4} to \ang{26}. Similarly, \cite{scolini2023} estimated the coherence scale of CMEs simulated within the EUHFORIA model and they found up to \ang{45} for the magnetic field strength.

In this study, we find significant changes on much smaller scales than the coherence scales computed in these studies. In fact, the scales described in this paper are almost absent in any CME model that relies on simulations, observations or theory. For instance, currently existing numerical model of CME propagation up to 1~au, \ang{2} typically represents a few grid cells \citep[\eg][]{vanderholst2014,pomoell2018,torok2018,liu2022a,regnault2023a} even when using non-uniform grids or adaptive mesh refinement techniques (or both) to increase the spatial resolution only where it is needed. 
 
To conclude, this study suggests that current CME models cannot explain discrepancies in the properties of a MC measured by two spacecraft even though they are separated by an angular separation as small as \ang{2}. The close proximity of the spacecraft along with the speed of fast magnetosonic waves allows us to rule out time evolution as being the main contributor of the observed discrepancies. 
The lack of sufficiently advanced models (both numerical and theoretical) to describe such small spatial scales can be explained by the limitation of numerical resources (having a higher spatial resolution would be too costly) but also by the paucity of CME observations like the one presented in this study, from which there is lot to learn.

\begin{acknowledgments}
F.~R., N.~A. and W.~Y. acknowledge grants 80NSSC21K0463  and 
AGS1954983. F.~R. and N.~A. acknowledge grant 80NSSC22K0349. N.~L. and F.~R. acknowledge grant 80NSSC20K0700. B.~Z. acknowledges grants 80NSSC23K1057 and AGS-2301382. C.~J.~F. acknowledges support from grants 80NSSC21K0463, and \textit{Wind} grant 80NSSC19K1293. E.~D. acknowledges funding by the European Union (ERC, HELIO4CAST, 101042188). Views and opinions expressed are however those of the author(s) only and do not necessarily reflect those of the European Union or the European Research Council Executive Agency. Neither the European Union nor the granting authority can be held responsible for them.
\end{acknowledgments}

\appendix

\section{Assessing the Calibration between SolO and \textit{Wind} measurements}
\label{sec:cross-correl}
\begin{figure}[ht!]
    \centering
    \includegraphics[width=.8\textwidth]{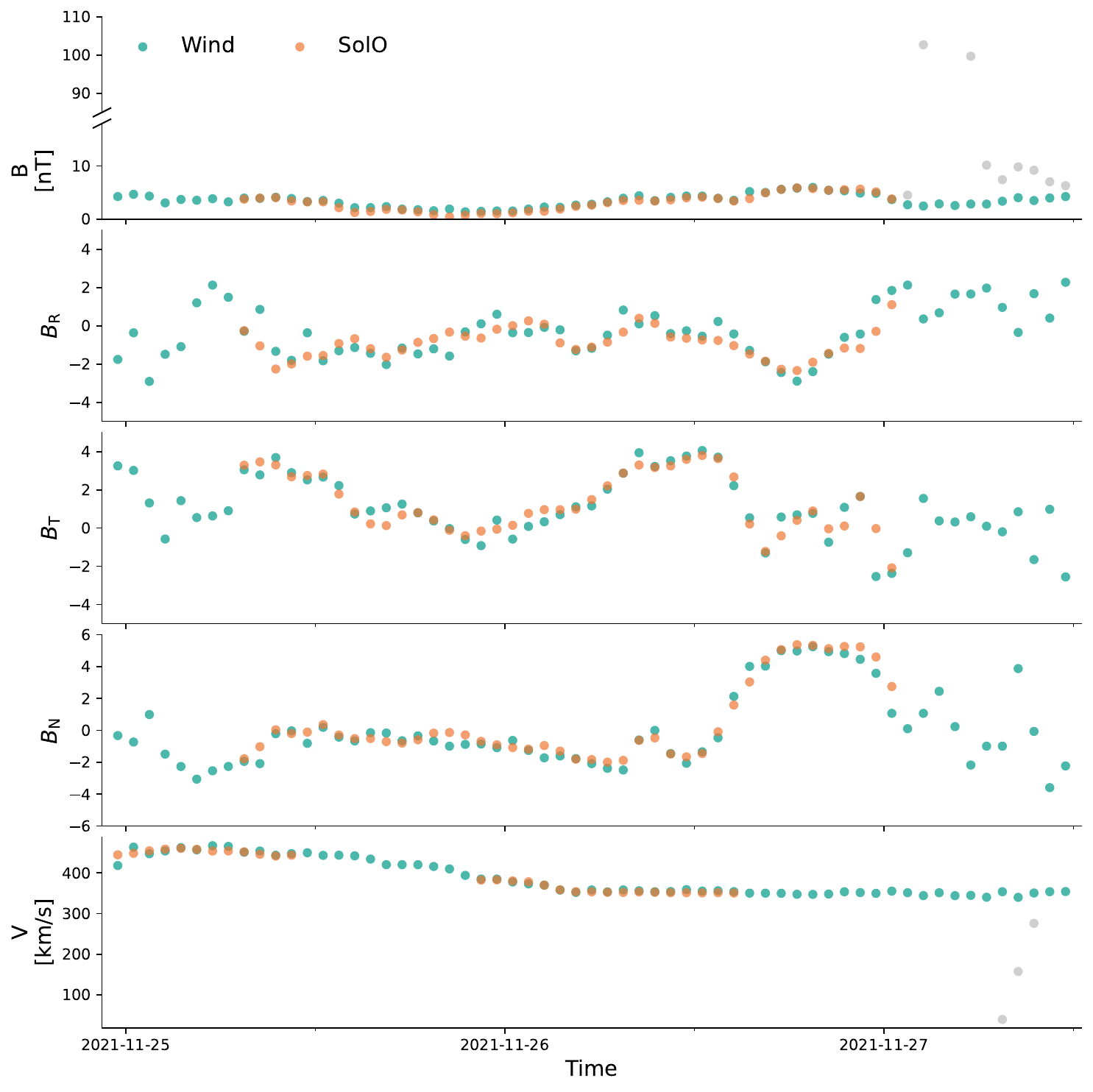}
    \caption{Cross correlation of the \textit{Wind} (green) and SolO (orange) hourly binned time series for the magnetic field strength (top panel) and speed (bottom panel) during SolO flyby close to the Earth. Grey points correspond to data not taken into account in the cross-correlation.}
    \label{fig:cross-correl}
\end{figure}

In this section, we compare the measurements made by \textit{Wind} and SolO  data during the SolO flyby close to the Earth ($< 0.01$ au) late November 2021. Figure~\ref{fig:cross-correl} shows the magnetic field (top panel), its RTN components (2nd, 3rd and 4th panel) and the speed (bottom panel) at \textit{Wind} (green) and at SolO (orange) as a function of time during the flyby. We can see that early on November, 27th 2021 the magnetic field quickly goes up to $\sim 100$ nT and the speed go as low as $\sim 40$ \kmps\ in the SolO measurements. This probably corresponds to SolO crossing Earth's magnetospheric boundaries as these values are not usually reached in normal observation of the solar wind properties close to 1 au. Thus, in order to compare \textit{Wind} and SolO measurements and not be affected by these artifacts in our calibration check, we remove the data in a 14h time window around it. Data points within this time window are shown in gray in Figure~\ref{fig:cross-correl}.

Table \ref{tab:cross-calib} shows the mean, median and maximum absolute difference between the magnetic field strength (and its components) and the speed measured by SolO and \Wind\ along with the standard deviation of the absolute difference.
Computing the absolute difference in the magnetic field strength and speed data between \textit{Wind} and SolO, we find a mean of $0.40$~nT and $5.2$~\kmps\ with a standard deviation of $0.34$~nT and $5.42$~\kmps\ and a maximum difference of $1.48$~nT and $26.37$~\kmps\ for the magnetic field strength and the speed respectively. 

\begin{table}[h!]
    \centering
    \begin{tabular}{c|c|c|c|c}
     Parameter & Mean & Median & Max & $\sigma$ \\
     \hline
    $B_{\rm mag}$ [nT] & 0.40 & 0.30 & 1.48 & 0.34 \\
    \Br\ [nT] & 0.52 & 0.40 &1.91 &0.43 \\
    \Bt\  [nT]& 0.41 & 0.28& 2.51 &0.43\\
    \Bn\  [nT]& 0.41 & 0.37 & 1.68 & 0.34 \\ 
    $V_{\rm mag}$ [\kmps] & 5.18 & 3.59 & 26.37 & 5.42\\
    \end{tabular}
    \caption{Mean and Median, Maximum absolute error and standard deviation of the difference between SolO and \Wind\ measurement when SolO was less than 0.01 au away from the Earth.}
    \label{tab:cross-calib}.
\end{table}

Similarly, we compute the median of the difference and find 0.30~nT and 3.59~\kmps. These values show that there is a good agreement between \textit{Wind} and SolO measurements for small separations. We also want to point out that, even if the separation between the two spacecraft is low, it is still non-zero. 
Assuming that the magnetic field of the solar wind decreases as $r^{\gamma}$ with $\gamma = 1.8 - 1.9$, 0.01 au is enough to see a decrease of $0.1$ nT for a magnetic field of $5$ nT, which probably contribute to the $0.4$ nT average difference of the magnetic field measurements. 

\bibliography{Research.bib}{}

\begin{thebibliography}{}
\expandafter\ifx\csname natexlab\endcsname\relax\def\natexlab#1{#1}\fi
\providecommand{\url}[1]{\href{#1}{#1}}
\providecommand{\dodoi}[1]{doi:~\href{http://doi.org/#1}{\nolinkurl{#1}}}
\providecommand{\doeprint}[1]{\href{http://ascl.net/#1}{\nolinkurl{http://ascl.net/#1}}}
\providecommand{\doarXiv}[1]{\href{https://arxiv.org/abs/#1}{\nolinkurl{https://arxiv.org/abs/#1}}}

\bibitem[{{Al-Haddad} {et~al.}(2022){Al-Haddad}, Galvin, Lugaz, Farrugia, \&
  Yu}]{al-haddad2022}
{Al-Haddad}, N., Galvin, A.~B., Lugaz, N., Farrugia, C.~J., \& Yu, W. 2022, The
  Astrophysical Journal, 927, 68, \dodoi{10.3847/1538-4357/ac32e1}

\bibitem[{Burlaga {et~al.}(1981)Burlaga, Sittler, Mariani, \&
  Schwenn}]{burlaga1981}
Burlaga, L., Sittler, E., Mariani, F., \& Schwenn, R. 1981, Journal of
  Geophysical Research: Space Physics, 86, 6673,
  \dodoi{10.1029/JA086iA08p06673}

\bibitem[{Burlaga \& Behannon(1982)}]{burlaga1982}
Burlaga, L.~F., \& Behannon, K.~W. 1982, Solar Physics, 81, 181,
  \dodoi{10.1007/BF00151989}

\bibitem[{Burt \& Smith(2012)}]{burt2012}
Burt, J., \& Smith, B. 2012, in 2012 {{IEEE Aerospace Conference}}, 1--13,
  \dodoi{10.1109/AERO.2012.6187025}

\bibitem[{Chiu {et~al.}(1998)Chiu, {Von-Mehlem}, Willey, Betenbaugh, Maynard,
  Krein, Conde, Gray, Hunt, Mosher, McCullough, Panneton, Staiger, \&
  Rodberg}]{chiu1998}
Chiu, M., {Von-Mehlem}, U., Willey, C., {et~al.} 1998, Space Science Reviews,
  86, 257, \dodoi{10.1023/A:1005002013459}

\bibitem[{Davies {et~al.}(2020)Davies, Forsyth, Good, \& Kilpua}]{davies2020}
Davies, E.~E., Forsyth, R.~J., Good, S.~W., \& Kilpua, E. K.~J. 2020, Solar
  Physics, 295, 157, \dodoi{10.1007/s11207-020-01714-z}

\bibitem[{Davies {et~al.}(2021{\natexlab{a}})Davies, Forsyth, Winslow,
  M{\"o}stl, \& Lugaz}]{davies2021b}
Davies, E.~E., Forsyth, R.~J., Winslow, R.~M., M{\"o}stl, C., \& Lugaz, N.
  2021{\natexlab{a}}, The Astrophysical Journal, 923, 136,
  \dodoi{10.3847/1538-4357/ac2ccb}

\bibitem[{Davies {et~al.}(2021{\natexlab{b}})Davies, M{\"o}stl, Owens, Weiss,
  Amerstorfer, Hinterreiter, Bauer, Bailey, Reiss, Forsyth, Horbury, O'Brien,
  Evans, Angelini, Heyner, Richter, Auster, Magnes, Baumjohann, Fischer,
  Barnes, Davies, \& Harrison}]{davies2021a}
Davies, E.~E., M{\"o}stl, C., Owens, M.~J., {et~al.} 2021{\natexlab{b}},
  Astronomy \& Astrophysics, 656, A2, \dodoi{10.1051/0004-6361/202040113}

\bibitem[{De~Lucas {et~al.}(2011)De~Lucas, Dal~Lago, Schwenn, \& Cl{\'u}a
  De~Gonzalez}]{delucas2011}
De~Lucas, A., Dal~Lago, A., Schwenn, R., \& Cl{\'u}a De~Gonzalez, A. 2011,
  Journal of Atmospheric and Solar-Terrestrial Physics, 73, 1361,
  \dodoi{10.1016/j.jastp.2011.02.007}

\bibitem[{Farrugia {et~al.}(1993)Farrugia, Burlaga, Osherovich, Richardson,
  Freeman, Lepping, \& Lazarus}]{farrugia1993}
Farrugia, C.~J., Burlaga, L.~F., Osherovich, V.~A., {et~al.} 1993, Journal of
  Geophysical Research: Space Physics, 98, 7621, \dodoi{10.1029/92JA02349}

\bibitem[{Fox {et~al.}(2016)Fox, Velli, Bale, Decker, Driesman, Howard, Kasper,
  Kinnison, Kusterer, Lario, Lockwood, McComas, Raouafi, \& Szabo}]{fox2016}
Fox, N.~J., Velli, M.~C., Bale, S.~D., {et~al.} 2016, Space Science Reviews,
  204, 7, \dodoi{10.1007/s11214-015-0211-6}

\bibitem[{Gonzalez \& Tsurutani(1987)}]{Gonzalez1987}
Gonzalez, W.~D., \& Tsurutani, B.~T. 1987, Planet. Space Sci., 35, 1101,
  \dodoi{10.1016/0032-0633(87)90015-8}

\bibitem[{Good \& Forsyth(2016)}]{good2016}
Good, S.~W., \& Forsyth, R.~J. 2016, Solar Physics, 291, 239,
  \dodoi{10.1007/s11207-015-0828-3}

\bibitem[{Good {et~al.}(2019)Good, Kilpua, LaMoury, Forsyth, Eastwood, \&
  M{\"o}stl}]{good2019}
Good, S.~W., Kilpua, E. K.~J., LaMoury, A.~T., {et~al.} 2019, Journal of
  Geophysical Research: Space Physics, 124, 4960, \dodoi{10.1029/2019JA026475}

\bibitem[{Gopalswamy {et~al.}(2001)Gopalswamy, Lara, Yashiro, Kaiser, \&
  Howard}]{gopalswamy2001}
Gopalswamy, N., Lara, A., Yashiro, S., Kaiser, M.~L., \& Howard, R.~A. 2001,
  Journal of Geophysical Research: Space Physics, 106, 29207,
  \dodoi{10.1029/2001JA000177}

\bibitem[{Gopalswamy {et~al.}(2009)Gopalswamy, M{\"a}kel{\"a}, Xie, Akiyama, \&
  Yashiro}]{gopalswamy2009b}
Gopalswamy, N., M{\"a}kel{\"a}, P., Xie, H., Akiyama, S., \& Yashiro, S. 2009,
  Journal of Geophysical Research: Space Physics, 114,
  \dodoi{10.1029/2008JA013686}

\bibitem[{Gopalswamy {et~al.}(2015)Gopalswamy, Yashiro, Xie, Akiyama, \&
  M{\"a}kel{\"a}}]{gopalswamy2015}
Gopalswamy, N., Yashiro, S., Xie, H., Akiyama, S., \& M{\"a}kel{\"a}, P. 2015,
  Journal of Geophysical Research: Space Physics, 120, 9221,
  \dodoi{10.1002/2015JA021446}

\bibitem[{Gulisano {et~al.}(2010)Gulisano, D{\'e}moulin, Dasso, Ruiz, \&
  Marsch}]{gulisano2010}
Gulisano, A.~M., D{\'e}moulin, P., Dasso, S., Ruiz, M.~E., \& Marsch, E. 2010,
  Astronomy and Astrophysics, 509, A39, \dodoi{10.1051/0004-6361/200912375}

\bibitem[{Horbury {et~al.}(2020)Horbury, O'Brien, Carrasco~Blazquez, Bendyk,
  Brown, Hudson, Evans, Oddy, Carr, Beek, Cupido, Bhattacharya, Dominguez,
  Matthews, Myklebust, Whiteside, Bale, Baumjohann, Burgess, Carbone, Cargill,
  Eastwood, Erd{\"o}s, Fletcher, Forsyth, Giacalone, Glassmeier, Goldstein,
  Hoeksema, Lockwood, Magnes, Maksimovic, Marsch, Matthaeus, Murphy,
  Nakariakov, Owen, Owens, {Rodriguez-Pacheco}, Richter, Riley, Russell,
  Schwartz, Vainio, Velli, Vennerstrom, Walsh, {Wimmer-Schweingruber}, Zank,
  M{\"u}ller, Zouganelis, \& Walsh}]{horbury2020}
Horbury, T.~S., O'Brien, H., Carrasco~Blazquez, I., {et~al.} 2020, Astronomy \&
  Astrophysics, 642, A9, \dodoi{10.1051/0004-6361/201937257}

\bibitem[{Kaiser \& Adams(2007)}]{kaiser2007}
Kaiser, M.~L., \& Adams, W.~J. 2007, in 2007 {{IEEE Aerospace Conference}},
  1--8, \dodoi{10.1109/AERO.2007.352745}

\bibitem[{Kaiser {et~al.}(2008)Kaiser, Kucera, Davila, St.~Cyr, Guhathakurta,
  \& Christian}]{kaiser2008}
Kaiser, M.~L., Kucera, T.~A., Davila, J.~M., {et~al.} 2008, Space Science
  Reviews, 136, 5, \dodoi{10.1007/s11214-007-9277-0}

\bibitem[{Kilpua {et~al.}(2021)Kilpua, Good, Dresing, Vainio, Davies, Forsyth,
  Gieseler, Lavraud, Asvestari, Morosan, Pomoell, Price, Heyner, Horbury,
  Angelini, O'Brien, Evans, {Rodriguez-Pacheco}, Herrero, Ho, \&
  {Wimmer-Schweingruber}}]{kilpua2021a}
Kilpua, E. K.~J., Good, S.~W., Dresing, N., {et~al.} 2021, Astronomy \&
  Astrophysics, 656, A8, \dodoi{10.1051/0004-6361/202140838}

\bibitem[{Leitner {et~al.}(2007)Leitner, Farrugia, M{\"o}stl, Ogilvie, Galvin,
  Schwenn, \& Biernat}]{leitner2007}
Leitner, M., Farrugia, C.~J., M{\"o}stl, C., {et~al.} 2007, Journal of
  Geophysical Research: Space Physics, 112, \dodoi{10.1029/2006JA011940}

\bibitem[{Lepping {et~al.}(1995)Lepping, Ac{\~u}na, Burlaga, Farrell, Slavin,
  Schatten, Mariani, Ness, Neubauer, Whang, Byrnes, Kennon, Panetta, Scheifele,
  \& Worley}]{lepping1995}
Lepping, R.~P., Ac{\~u}na, M.~H., Burlaga, L.~F., {et~al.} 1995, Space Science
  Reviews, 71, 207, \dodoi{10.1007/BF00751330}

\bibitem[{Li {et~al.}(2022)Li, Wang, Guo, \& Lyu}]{li2022}
Li, X., Wang, Y., Guo, J., \& Lyu, S. 2022, The Astrophysical Journal Letters,
  928, L6, \dodoi{10.3847/2041-8213/ac5b72}

\bibitem[{Liu {et~al.}(2005)Liu, Richardson, \& Belcher}]{liu2005}
Liu, Y., Richardson, J., \& Belcher, J. 2005, Planetary and Space Science, 53,
  3, \dodoi{10.1016/j.pss.2004.09.023}

\bibitem[{Liu {et~al.}(2022)Liu, Shen, Yang, \& Ma}]{liu2022a}
Liu, Y., Shen, F., Yang, Y., \& Ma, M. 2022, The Astrophysical Journal, 940,
  11, \dodoi{10.3847/1538-4357/ac9b16}

\bibitem[{Lugaz {et~al.}(2018)Lugaz, Farrugia, Winslow, {Al-Haddad}, Galvin,
  {Nieves-Chinchilla}, Lee, \& Janvier}]{lugaz2018}
Lugaz, N., Farrugia, C.~J., Winslow, R.~M., {et~al.} 2018, The Astrophysical
  Journal, 864, L7, \dodoi{10.3847/2041-8213/aad9f4}

\bibitem[{Lugaz {et~al.}(2020)Lugaz, Salman, Winslow, {Al-Haddad}, Farrugia,
  Zhuang, \& Galvin}]{lugaz2020a}
Lugaz, N., Salman, T.~M., Winslow, R.~M., {et~al.} 2020, The Astrophysical
  Journal, 899, 119, \dodoi{10.3847/1538-4357/aba26b}

\bibitem[{Lugaz {et~al.}(2022)Lugaz, Salman, Zhuang, {Al-Haddad}, Scolini,
  Farrugia, Yu, Winslow, M{\"o}stl, Davies, \& Galvin}]{lugaz2022}
Lugaz, N., Salman, T.~M., Zhuang, B., {et~al.} 2022, The Astrophysical Journal,
  929, 149, \dodoi{10.3847/1538-4357/ac602f}

\bibitem[{M{\"o}stl {et~al.}(2022)M{\"o}stl, Weiss, Reiss, Amerstorfer, Bailey,
  Hinterreiter, Bauer, Barnes, Davies, Harrison, {Freiherr von Forstner},
  Davies, Heyner, Horbury, \& Bale}]{mostl2022}
M{\"o}stl, C., Weiss, A.~J., Reiss, M.~A., {et~al.} 2022, The Astrophysical
  Journal Letters, 924, L6, \dodoi{10.3847/2041-8213/ac42d0}

\bibitem[{M{\"u}ller {et~al.}(2020)M{\"u}ller, St.~Cyr, Zouganelis, Gilbert,
  Marsden, {Nieves-Chinchilla}, Antonucci, Auch{\`e}re, Berghmans, Horbury,
  Howard, Krucker, Maksimovic, Owen, Rochus, {Rodriguez-Pacheco}, Romoli,
  Solanki, Bruno, Carlsson, Fludra, Harra, Hassler, Livi, Louarn, Peter,
  Sch{\"u}hle, Teriaca, {del Toro Iniesta}, {Wimmer-Schweingruber}, Marsch,
  Velli, De~Groof, Walsh, \& Williams}]{muller2020}
M{\"u}ller, D., St.~Cyr, O.~C., Zouganelis, I., {et~al.} 2020, Astronomy \&
  Astrophysics, 642, A1, \dodoi{10.1051/0004-6361/202038467}

\bibitem[{Ogilvie {et~al.}(1995)Ogilvie, Chornay, Fritzenreiter, Hunsaker,
  Keller, Lobell, Miller, Scudder, Sittler, Torbert, Bodet, Needell, Lazarus,
  Steinberg, Tappan, Mavretic, \& Gergin}]{ogilvie1995}
Ogilvie, K.~W., Chornay, D.~J., Fritzenreiter, R.~J., {et~al.} 1995, Space
  Science Reviews, 71, 55, \dodoi{10.1007/BF00751326}

\bibitem[{Owen {et~al.}(2020)Owen, Bruno, Livi, Louarn, Al~Janabi, Allegrini,
  Amoros, Baruah, Barthe, Berthomier, Bordon, {Brockley-Blatt}, Brysbaert,
  Capuano, Collier, DeMarco, Fedorov, Ford, Fortunato, Fratter, Galvin,
  Hancock, Heirtzler, Kataria, Kistler, Lepri, Lewis, Loeffler, Marty, Mathon,
  Mayall, Mele, Ogasawara, Orlandi, Pacros, Penou, Persyn, Petiot, Phillips,
  P{\v r}ech, Raines, Reden, Rouillard, Rousseau, Rubiella, Seran, Spencer,
  Thomas, Trevino, Verscharen, Wurz, Alapide, Amoruso, Andr{\'e}, Anekallu,
  Arciuli, Arnett, Ascolese, Bancroft, Bland, Brysch, Calvanese, Castronuovo,
  {\v C}erm{\'a}k, Chornay, Clemens, Coker, Collinson, D'Amicis, Dandouras,
  Darnley, Davies, Davison, De~Los~Santos, Devoto, Dirks, Edlund, Fazakerley,
  Ferris, Frost, Fruit, Garat, G{\'e}not, Gibson, Gilbert, {de Giosa}, Gradone,
  Hailey, Horbury, Hunt, Jacquey, Johnson, Lavraud, Lawrenson, Leblanc,
  Lockhart, Maksimovic, Malpus, Marcucci, Mazelle, Monti, Myers, Nguyen,
  {Rodriguez-Pacheco}, Phillips, Popecki, Rees, Rogacki, Ruane, Rust, Salatti,
  Sauvaud, Stakhiv, Stange, Stubbs, Taylor, Techer, Terrier, Thibodeaux,
  Urdiales, Varsani, Walsh, Watson, Wheeler, Willis, {Wimmer-Schweingruber},
  Winter, Yardley, \& Zouganelis}]{owen2020}
Owen, C.~J., Bruno, R., Livi, S., {et~al.} 2020, Astronomy \& Astrophysics,
  642, A16, \dodoi{10.1051/0004-6361/201937259}

\bibitem[{Owens(2009)}]{owens2009}
Owens, M.~J. 2009, Solar Physics, 260, 207, \dodoi{10.1007/s11207-009-9442-6}

\bibitem[{Owens {et~al.}(2005)Owens, Cargill, Pagel, Siscoe, \&
  Crooker}]{owens2005}
Owens, M.~J., Cargill, P.~J., Pagel, C., Siscoe, G.~L., \& Crooker, N.~U. 2005,
  Journal of Geophysical Research: Space Physics, 110,
  \dodoi{10.1029/2004JA010814}

\bibitem[{Pomoell \& Poedts(2018)}]{pomoell2018}
Pomoell, J., \& Poedts, S. 2018, Journal of Space Weather and Space Climate, 8,
  A35, \dodoi{10.1051/swsc/2018020}

\bibitem[{Raouafi {et~al.}(2023)Raouafi, Matteini, Squire, Badman, Velli,
  Klein, Chen, Matthaeus, Szabo, Linton, Allen, Szalay, Bruno, Decker,
  {Akhavan-Tafti}, Agapitov, Bale, Bandyopadhyay, Battams, Ber{\v c}i{\v c},
  Bourouaine, Bowen, Cattell, Chandran, Chhiber, Cohen, D'Amicis, Giacalone,
  Hess, Howard, Horbury, Jagarlamudi, Joyce, Kasper, Kinnison, Laker, Liewer,
  Malaspina, Mann, McComas, {Niembro-Hernandez}, {Nieves-Chinchilla},
  Panasenco, Pokorn{\'y}, Pusack, Pulupa, Perez, Riley, Rouillard, Shi,
  Stenborg, Tenerani, Verniero, Viall, Vourlidas, Wood, Woodham, \&
  Woolley}]{raouafi2023}
Raouafi, N.~E., Matteini, L., Squire, J., {et~al.} 2023, Space Science Reviews,
  219, 8, \dodoi{10.1007/s11214-023-00952-4}

\bibitem[{Regnault {et~al.}(2023{\natexlab{a}})Regnault, {Al-Haddad}, Lugaz,
  Farrugia, Yu, Davies, Galvin, \& Zhuang}]{regnault2023b}
Regnault, F., {Al-Haddad}, N., Lugaz, N., {et~al.} 2023{\natexlab{a}}, Accepted
  for publication at The Astrophysical Journal

\bibitem[{Regnault {et~al.}(2023{\natexlab{b}})Regnault, Strugarek, Janvier,
  Auch{\`e}re, Lugaz, \& {Al-Haddad}}]{regnault2023a}
Regnault, F., Strugarek, A., Janvier, M., {et~al.} 2023{\natexlab{b}},
  Astronomy \& Astrophysics, 670, A14, \dodoi{10.1051/0004-6361/202244483}

\bibitem[{Richardson \& Cane(1995)}]{richardson1995}
Richardson, I.~G., \& Cane, H.~V. 1995, Journal of Geophysical Research, 100,
  23397, \dodoi{10.1029/95JA02684}

\bibitem[{Riley {et~al.}(2018)Riley, Mays, Andries, Amerstorfer, Biesecker,
  Delouille, Dumbovi{\'c}, Feng, Henley, Linker, M{\"o}stl, Nu{\~n}ez, Pizzo,
  Temmer, Tobiska, Verbeke, West, \& Zhao}]{riley2018}
Riley, P., Mays, M.~L., Andries, J., {et~al.} 2018, Space Weather, 16, 1245,
  \dodoi{10.1029/2018SW001962}

\bibitem[{Salman {et~al.}(2020)Salman, Winslow, \& Lugaz}]{salman2020a}
Salman, T.~M., Winslow, R.~M., \& Lugaz, N. 2020, Journal of Geophysical
  Research: Space Physics, 125, e2019JA027084, \dodoi{10.1029/2019JA027084}

\bibitem[{Scolini {et~al.}(2021)Scolini, Winslow, Lugaz, \&
  Poedts}]{scolini2021g}
Scolini, C., Winslow, R.~M., Lugaz, N., \& Poedts, S. 2021, The Astrophysical
  Journal Letters, 916, L15, \dodoi{10.3847/2041-8213/ac0d58}

\bibitem[{Scolini {et~al.}(2023)Scolini, Winslow, Lugaz, \&
  Poedts}]{scolini2023}
---. 2023, The Astrophysical Journal, 944, 46, \dodoi{10.3847/1538-4357/aca893}

\bibitem[{Thernisien(2011)}]{thernisien2011}
Thernisien, A. 2011, The Astrophysical Journal Supplement Series, 194, 33,
  \dodoi{10.1088/0067-0049/194/2/33}

\bibitem[{T{\"o}r{\"o}k {et~al.}(2018)T{\"o}r{\"o}k, Downs, Linker, Lionello,
  Titov, Miki{\'c}, Riley, Caplan, \& Wijaya}]{torok2018}
T{\"o}r{\"o}k, T., Downs, C., Linker, J.~A., {et~al.} 2018, The Astrophysical
  Journal, 856, 75, \dodoi{10.3847/1538-4357/aab36d}

\bibitem[{Trotta {et~al.}(2023)Trotta, Hietala, Horbury, Dresing, Vainio,
  Wilson, Plotnikov, \& Kilpua}]{trotta2023}
Trotta, D., Hietala, H., Horbury, T., {et~al.} 2023, Monthly Notices of the
  Royal Astronomical Society, 520, 437, \dodoi{10.1093/mnras/stad104}

\bibitem[{{van der Holst} {et~al.}(2014){van der Holst}, Sokolov, Meng, Jin,
  Manchester, T{\'o}th, \& Gombosi}]{vanderholst2014}
{van der Holst}, B., Sokolov, I.~V., Meng, X., {et~al.} 2014, The Astrophysical
  Journal, 782, 81, \dodoi{10.1088/0004-637X/782/2/81}

\bibitem[{Vr{\v s}nak {et~al.}(2013)Vr{\v s}nak, {\v Z}ic, Vrbanec, Temmer,
  Rollett, M{\"o}stl, Veronig, {\v C}alogovi{\'c}, Dumbovi{\'c}, Luli{\'c},
  Moon, \& Shanmugaraju}]{vrsnak2013}
Vr{\v s}nak, B., {\v Z}ic, T., Vrbanec, D., {et~al.} 2013, Solar Physics, 285,
  295, \dodoi{10.1007/s11207-012-0035-4}

\bibitem[{Wang {et~al.}(2005)Wang, Du, \& Richardson}]{wang2005}
Wang, C., Du, D., \& Richardson, J.~D. 2005, Journal of Geophysical Research:
  Space Physics, 110, \dodoi{10.1029/2005JA011198}

\bibitem[{Wang {et~al.}(2011)Wang, Chen, Gui, Shen, Ye, \& Wang}]{wang2011}
Wang, Y., Chen, C., Gui, B., {et~al.} 2011, Journal of Geophysical Research:
  Space Physics, 116, \dodoi{10.1029/2010JA016101}

\bibitem[{Winslow {et~al.}(2015)Winslow, Lugaz, Philpott, Schwadron, Farrugia,
  Anderson, \& Smith}]{winslow2015}
Winslow, R.~M., Lugaz, N., Philpott, L.~C., {et~al.} 2015, Journal of
  Geophysical Research: Space Physics, 120, 6101, \dodoi{10.1002/2015JA021200}

\bibitem[{Winslow {et~al.}(2021)Winslow, Lugaz, Scolini, \&
  Galvin}]{winslow2021}
Winslow, R.~M., Lugaz, N., Scolini, C., \& Galvin, A.~B. 2021, The
  Astrophysical Journal, 916, 94, \dodoi{10.3847/1538-4357/ac0821}

\bibitem[{Wold {et~al.}(2018)Wold, Mays, Taktakishvili, Jian, Odstrcil, \&
  MacNeice}]{wold2018}
Wold, A.~M., Mays, M.~L., Taktakishvili, A., {et~al.} 2018, Journal of Space
  Weather and Space Climate, 8, A17, \dodoi{10.1051/swsc/2018005}

\bibitem[{Xu {et~al.}(2011)Xu, Wei, \& Feng}]{xu2011}
Xu, X., Wei, F., \& Feng, X. 2011, Journal of Geophysical Research: Space
  Physics, 116, \dodoi{10.1029/2010JA016159}

\end{thebibliography}
\bibliographystyle{aasjournal}

\end{document}